\documentclass[12pt]{iopart}
\usepackage{epsf}

\newcommand{\La}{Sr$_{0.9}$La$_{0.1}$CuO$_{2}$}
\newcommand{\Y}{YBa$_2$Cu$_3$O$_{7-\delta}$}
\newcommand{\YB}{YBa$_2$Cu$_4$O$_8$}

\begin{document}

\rapid{Comparative study of the pressure effects on the magnetic
penetration depth in electron- and hole-doped cuprate
superconductors}

\author{D. Di Castro$^1$\footnote[5]{e-mail:
daniele.di.castro@uniroma2.it}, R. Khasanov$^{2,3}$, A.
Shengelaya$^4$, K.~Conder$^5$, D.-J. Jang$^6$, M.-S. Park$^6$,
S.-I. Lee$^7$, and H.~Keller$^2$
 }

\address{$^1$ CNR-INFM-Coherentia and Dipartimento di Ingegneria Meccanica,
Universita' di Roma "Tor Vergata", Via del Politecnico 1, I-00133
Roma, Italy}

\address{$^2$ Physik-Institut der Universit\"{a}t Z\"{u}rich,
Winterthurerstrasse 190, CH-8057, Z\"urich, Switzerland}

\address{$^3$ Laboratory for Muon Spin Spectroscopy, Paul Scherrer
Institut, CH-5232 Villigen PSI, Switzerland}

\address{$^4$ Physics Institute of Tbilisi State University, Chavchavadze 3, GE-0128 Tbilisi,
Georgia}

\address{$^5$ Laboratory for Developments and Methods, Paul Scherrer Institute,
CH-5232 Villigen PSI, Switzerland}

\address{$^6$ National Creative Research Initiative Center for
Superconductivity and Department of Physics, Pohang University of
Science and Technology, Pohang 790-784, Republic of Korea}

\address{$^7$ National Creative Research Initiative Center for Superconductivity,
Department of Physics, Sogang University, Seoul, Korea}

%\author{D. Di Castro}
%%\affiliation{Physik-Institut der Universit\"{a}t Z\"{u}rich, Winterthurerstrasse 190, CH-8057 Z\"urich,
%Switzerland} %%
%\affiliation{CNR-INFM-Coherentia and Dipartimento di Fisica,
%Universita' di Roma "La Sapienza", P.le A. Moro 2, I-00185 Roma,
%Italy} %%
%
%author{R. Khasanov}
%\affiliation{Physik-Institut der Universit\"{a}t Z\"{u}rich,
%Winterthurerstrasse 190, CH-8057 Z\"urich, Switzerland} %%
%\affiliation{Laboratory for Muon Spin Spectroscopy, Paul Scherrer
%Institut, CH-5232 Villigen PSI, Switzerland} %%
%author{A. Shengelaya}
%\affiliation{Physics Institute of Tbilisi State University, Chavchavadze 3, GE-0128 Tbilisi, Georgia}
%%
%\author{K.~Conder}
% \affiliation{}
%%
%\author{C. Grimaldi}
%% \affiliation{LPM, Ecole Polytechnique Fédérale de Lausanne, Station 17, CH-1015 Lausanne, Switzerland}
%

%\author{D.-J. Jang}
%\affiliation{National Creative Research Initiative Center for
%uperconductivity and Department of Physics,
%Pohang University of Science and Technology, Pohang 790-784, Republic of Korea } %
%%
%\author{M.-S. Park}
%\affiliation{National Creative Research Initiative Center for
%Superconductivity and Department of Physics,
%ohang University of Science and Technology, Pohang 790-784, Republic of Korea } %,
%%
%\author{S.-I. Lee}
%\affiliation{National Creative Research Initiative Center for
%uperconductivity and Department of Physics,
%Pohang University of Science and Technology, Pohang 790-784, Republic of Korea } %
%%
%\author{H. Keller}
%\affiliation{Physik-Institut der Universit\"{a}t Z\"{u}rich,
%Winterthurerstrasse 190, CH-8057 Z\"urich, Switzerland}

\begin{abstract}
The effect of pressure on the magnetic penetration depth $\lambda$
was tested in the hole-doped superconductor \Y~and in the
electron-doped one \La~ by means of magnetization measurements.
Whereas a large change of $\lambda$
 was found in \Y, confirming the non-adiabatic character of the electron-phonon coupling in hole-doped superconductors,
 the same quantity  is not affected by  pressure in the electron-doped \La, suggesting a close
   similarity of the latter to conventional adiabatic BCS superconductors. The present results imply a remarkable
  difference between  the electronic properties of hole-doped cuprates and the electron-doped \La, giving a strong
  contribution to the long debated asymmetric consequences of hole- and  electron-doping in cuprate superconductors.
\end{abstract}

%\pacs{74.72.-h, 74.62.Fj, 71.38.-k, 74.25.Ha}

\maketitle
High transition temperature ($T_c$) superconductivity in cuprates
is obtained by introducing holes or electrons into the
antiferromagnetic parent compound. Both electron ($n$-HTS's) and
hole  ($p$-HTS's) doped cuprate superconductors share a common
building block, i.e., the copper-oxygen plane. However, there are
a number of important differences between the generic phase
diagrams of the electron-doped and hole-doped materials, which are
commonly known  as the "electron-hole asymmetry" problem.   For
example, the doping ranges where the antiferromagnetic state and
the superconducting state emerge in $n$-HTS's are different from
those in $p$-HTS's, and the existence of a pseudogap in $n$-HTS's,
certainly present in $p$-HTS's, is still object of debates.
Recently, Shengelaya {\em et al.} \cite{Shengelaya} showed that
$n$-HTS's do not follow the Uemura relation \cite{Uemura89}
between $T_c$ and the superfluid density ($\lambda^{-2}$), found
for $p$-HTS's, indicating a remarkable difference between these
two families of superconductors.
 Moreover, although in $n$-HTS's a d-wave
  symmetry of the order parameter has been reported \cite{Tsuein,Prozorov,Chesca},
  however there are also strong experimental evidences which indicate
  for $n$-HTS's the
 existence of  conventional s-wave
 symmetry \cite{KimPRL03,Alff,ChenPRL02}.
 This is in apparent contrast with $p$-HTS's,
 where d-wave pairing symmetry is well accepted (see for example Refs.~\cite{VHarlingen,Tsuei}), although a
 multi-component (d+s-wave) order parameter is now
 acquiring overwhelming evidences \cite{KhasLa,Khas123}. These asymmetric behaviors raised the fundamental question
 whether or not the mechanism of superconductivity in $n$-HTS's is common to that one in $p$-HTS's.

An important characteristic feature of the $p$-HTS's is the
presence of a strong electron-phonon interaction, which leads to
non-adiabatic effects and polaron formation. Indeed, in $p$-HTS's,
induced lattice modification, by, e.g., oxygen isotope
substitution \cite{Zhao97,Hofer00,Khasanov03,Khasanov04}  or
application of external pressure \cite{KhasanovY124Pres}, led to
substantial changes in the superconducting critical temperature
$T_c$ and the magnetic penetration depth $\lambda(0)$. Since
$\lambda(0)$ is related to the effective mass $m^{\ast}$, these
results were interpreted in the framework of $non-adiabatic$
theory of the electron-phonon interaction \cite{sarkar,Grimaldi}
and of polaron superconductivity \cite{Alexandrov94,Keller}. The
conventional phonon-mediated theory of superconductivity is based
on the Migdal adiabatic approximation, in which  $m^{\ast}$ is
independent of the lattice vibrations. However, if the coupling
between the carriers and the lattice is strong enough, and the
typical phonon frequency $\omega_{ph}$ is comparable to the Fermi
energy $E_F$, the Migdal adiabatic approximation breaks down and
$m^{\ast}$ depends on the lattice degrees of freedom, with the
opening of new interaction channels which give rise to, {\it e.
g.}, anomalous pressure and isotope effects
\cite{sarkar,Grimaldi}.

Whereas non-adiabatic interaction appears to be a characteristic
feature of $p$-HTS's, on the contrary in low temperature BCS
superconductors  the  adiabatic approximation usually holds. For
example, the BCS low temperature superconductors RbOs$_2$O$_6$
\cite{KhasanovRbOs} and YB$_6$ \cite{KhasanovYB6}, whereas showing
a $T_c$ shift with pressure, do not present any pressure effect on
$\lambda(0)$, indicating the adiabatic character of the
electron-lattice interaction in these systems. A limit case is
MgB$_2$. Studies of the pressure \cite{DiCastroPress} and boron
isotope \cite{DiCastroBIE} effects
 evidenced shifts of $\lambda(0)$ compatible with the adiabatic
limit.

To  check whether the electron-hole asymmetry in HTS does concern
also the nature of the electron-phonon coupling, in this work we
measured the pressure effect on $\lambda$ in the $n$-HTS
Sr$_{0.9}$La$_{0.1}$CuO$_2$, by means of magnetization
measurements under pressure. This system belongs to the family of
electron doped infinite-layer superconductors (ILS's). This class
of materials has the simplest crystal structure among all cuprates
superconductors, and the charge reservoir block, commonly present
in cuprates, does not exist in the infinite-layer structure.
Moreover, the buckling of CuO$_2$ plane is absent
\cite{Jorgensen}, and the oxygen content is stoichiometric
without vacancies or interstitial oxygen \cite{Jorgensen}, which,
instead, is a common problem of other $n$-HTS's and $p$-HTS's
families. These properties allow to study the effect of pressure
on this system avoiding  modification of $n_s$ (superconducting
 carrier density) and $T_c$ via secondary route, as, for
example, charge transfer processes.

Finally,
 ILS's have much higher $T_c$ ($\simeq$ 43 K) compared to
the other $n$-HTS's. For comparison, we also report the same
measurement on the $p$-HTS YBa$_2$Cu$_3$O$_{7-\delta}$.
 The temperature
dependence of the inverse squared magnetic penetration depth
$\lambda^{-2}$ was extracted from  Meissner fraction measurements
at  low magnetic field. Very small and negligible pressure effects
on $T_c$ were found in YBa$_2$Cu$_3$O$_7$ and
Sr$_{0.9}$La$_{0.1}$CuO$_2$, respectively. Whereas a pronounced
pressure effect on $\lambda^{-2}$ was revealed in the $p$-HTS
YBa$_2$Cu$_3$O$_7$ at low temperature, zero pressure effect was
detected in the $n$-HTS Sr$_{0.9}$La$_{0.1}$CuO$_2$, suggesting
that this superconductor is in the adiabatic limit.

A high quality   polycrystalline sample of
Sr$_{0.9}$La$_{0.1}$CuO$_2$ with a sharp superconducting
transition $T_c$ $\simeq$ 43 K was synthesized by using a cubic
multianvil press \cite{Jung02}. The $p$-HTS polycrystalline sample
of YBa$_2$Cu$_3$O$_7$ was prepared by standard solid state
reaction \cite{Conder}. The samples, which are both close to
optimal doping,
 were mixed with
Fluorinert FC77 (pressure transmitting medium) with a sample to
liquid volume ratio of approximately 1/6. The pressure was
generated in a copper-berillium piston cylinder clamp, which
allows to reach hydrostatic pressures up to 1.2 GPa. The pressure
was measured in situ by monitoring the $T_c$ shift of a small
piece of Pb included in the pressure cell. The value of the
Meissner fraction was calculated from 0.5 mT field-cooled (FC)
SQUID magnetization measurements assuming spherical grains. The
temperature dependence of the effective (powder average)
penetration depth was calculated from the measured Meissner
fraction by using the Shoenberg model \cite{Shoenb}. For
anisotropic polycrystalline superconductors, the effective
penetration depth is dominated by the in plane contribution
($\lambda$ = 1.31 $\lambda_{ab}$ \cite{Fesenko}). Therefore, the
effective penetration depth evaluated in this study is mainly a
measure of the in plane penetration depth $\lambda_{ab}$. The
error bars on $\lambda^{-2}$ were determined by the
reproducibility in repeated measurements.
\begin{figure}[htb]
\begin{center}
%\begin{flushright}
\epsfxsize = 10cm \epsfbox{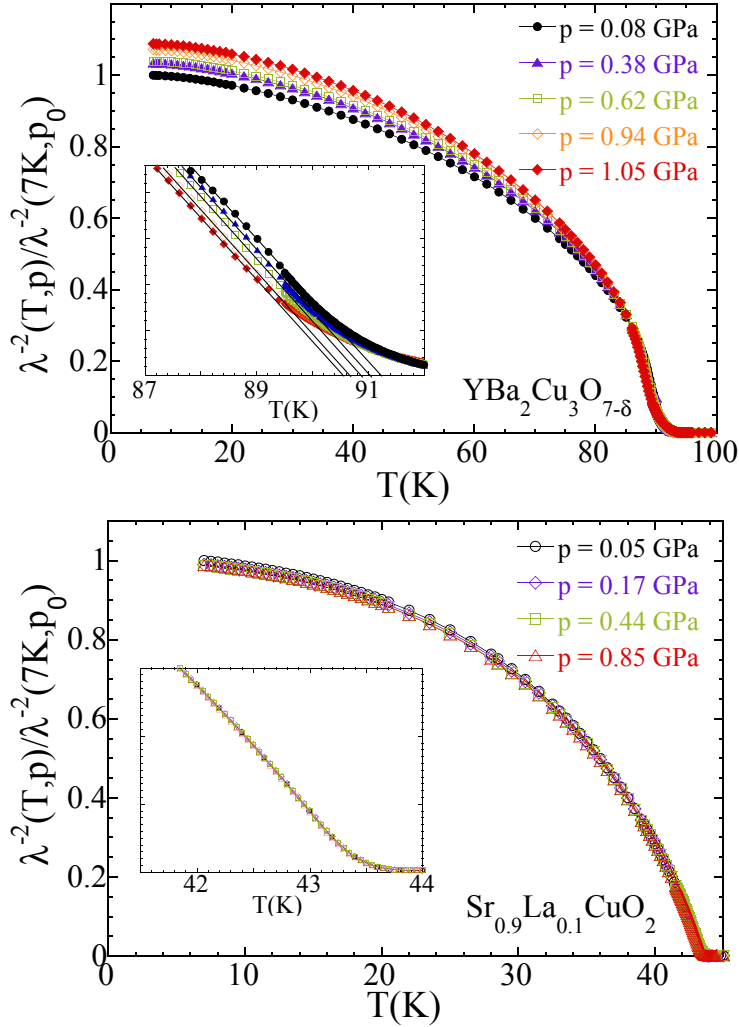}
\end{center}
%\end{flushright}
\vspace{-1cm}
 \caption{(color online) Temperature dependence of $\lambda^{-2}$
for \Y~ (upper panel) and \La~ (lower panel) at different
pressures. The insets show the same data for enlarged temperature
scale in the region close to $T_c$.}
\label{fig1}
\end{figure}

In Fig. 1, the temperature dependence of $\lambda^{-2}$ for \Y~
and \La~  at different pressures is shown. We note here that in
the present work we are not focusing on the temperature dependence
of $\lambda^{-2}$, which, in non-aligned polycrystalline powder,
can be  affected, especially at low temperature, by impurity
scattering \cite{Hirschfeld}, chemical and/or structural defects
\cite{Panagopoulos}, and  by the c-axis contribution, although the
latter is  small in an anisotropic superconductor  \cite{Fesenko}.
Here we are interested only in the relative shift of
$\lambda^{-2}(0)$ with pressure, which instead is not affected by
all the above contributions. Due to the unknown average grain
size, and thus the unknown absolute value of $\lambda$, the data
in Fig. 1 are normalized to the value of $\lambda^{-2}$ at the
lowest temperature, $T_m$ = 7 K, and pressure, p$_0$ (p$_0$ = 0.05
GPa for \La~ and 0.08 GPa for \Y). Data at temperatures lower than
$T_m$ are affected by the superconducting transition of Pb, used
to measure the pressure. A pronounced pressure effect on
$\lambda^{-2}$ is present in \Y~
 at low temperature, whereas no pressure effect is observed for  \La~  within  errors. Insets
of Fig. 1 show in details the region close to $T_c$ for the two
compounds. The \Y~ sample shows a small shift of the
$\lambda^{-2}$(T) curves with pressure, related to a corresponding
small decrease of the critical temperature. The $T_c$'s at
different pressure were estimated by a linear extrapolation to
$\lambda^{-2}$ = 0 (see the inset of Fig. 1). The results are
shown in the inset of Fig. 3. A linear fit gives $dT_c/dp$ =
-0.69(5) K/GPa. Different types of $p$-HTS's show various
pressure-induced effects on $T_c$, attributed to charge transfer,
constant shift in $T_c^{max}$ (where $T_c^{max}$ corresponds to
the optimally doped value), and to thermal activated oxygen
ordering (see, i.e., Ref. ~\cite{Fietz}). However, usually
$dT_c/dp$ peaks in the underdoped region of the phase diagram and
tends to zero near  optimal doping. In the case of optimal- and
over-doped \Y, the main contribution arises from the Cu-O chain to
the CuO$_2$ plane charge transfer \cite{Fietz}. It is not
surprising, therefore, that our \Y~ sample, which is close to
optimum doping, exhibits a very small and negative shift of $T_c$
with pressure.

\begin{figure}[htb]
\begin{center}
%\begin{flushright}
\epsfxsize = 15cm \epsfbox{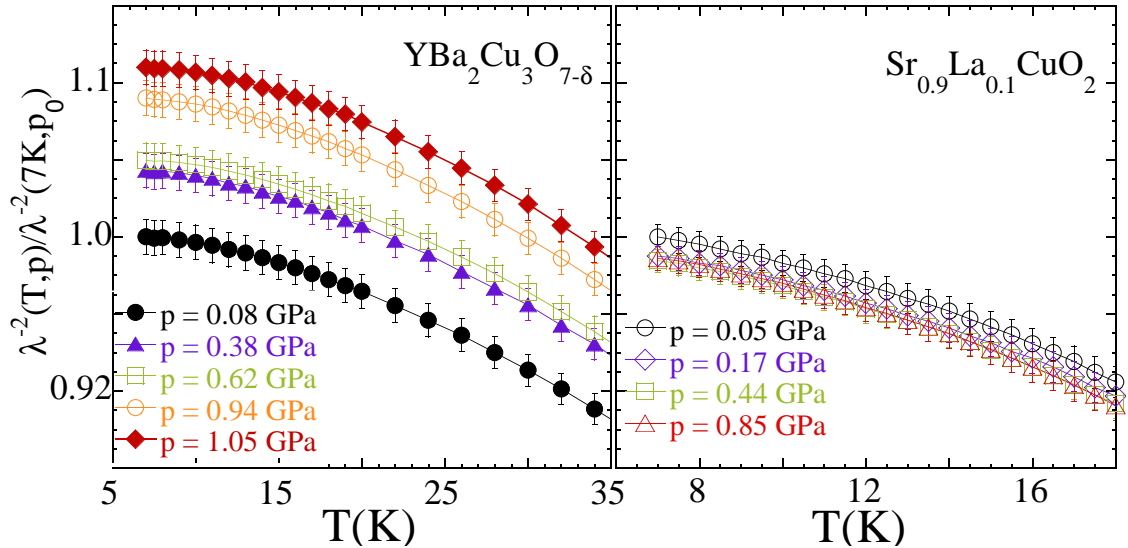}
\end{center}
%\end{flushright}
\vspace{-1cm}
 \caption{(color online) Temperature dependence of $\lambda^{-2}$
for \Y~ (left panel) and \La~ (right panel) at different pressures
in the low temperature region, shown on the same vertical axis
scale.}\label{fig2}
\end{figure}

In the case of the $n$-HTS \La, there is an almost complete
overlap of the curves at different pressures close to $T_c$ (inset
to the lower panel of Fig. 1), indicating absence of a pressure
effect on $T_c$ in this system. This result is in agreement with
previous reports. Indeed, the onset of superconductivity was found
to be almost pressure independent in some $n$-HTS's
\cite{Murayama,Crusellas,Bobrovskii} and, in particular, zero
pressure effect on $T_c$ was already previously found in
\La~\cite{Kim}. These findings have been attributed for example to
the absence of the apical oxygen in $n$-HTS \cite{Murayama}, or to
the fact that in \La~ $\xi_c$ is larger than the inter-CuO$_2$
layer distance and therefore  a further enhancement of the
inter-layer coupling by lattice compression should not enhance
superconductivity \cite{Kim}. Moreover, the absence of the charge
reservoir block makes pressure induced charge transfer to the Cu-O
planes unlikely \cite{Kim}.

%The present experiment thus confirms
%previous results.
% A very small negative pressure effect (-0.75(30) K/GPa) was measured by Bobrovskii
%{\em et al.} \cite{Bobrovskii} in a similar compound
%(Sr$_{0.9}$Pr$_{0.1}$CuO$_{2}$), which was tentatively  explained
%as due to the anisotropic lattice compression which transfers the
%system from the optimally doped regime to the overdoped one, where
%also $p$-HTS's show negative $dT_c/dp$.
%%
\begin{figure}[htb]
\begin{center}
%\begin{flushright}
\epsfxsize = 15cm \epsfbox{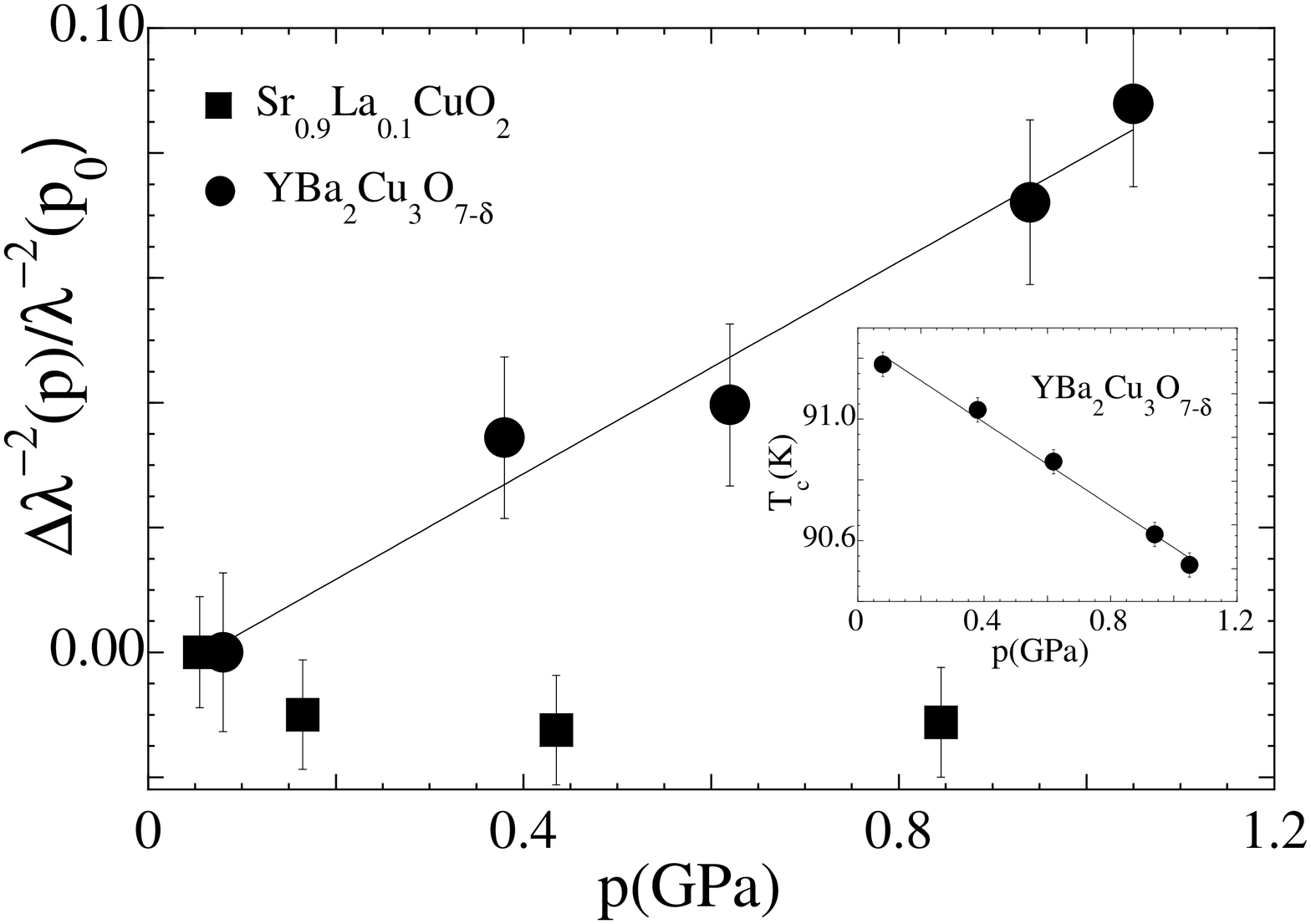}
\end{center}
%\end{flushright}
\vspace{-1cm}
 \caption{Pressure dependence of $\Delta \lambda^{-2}
/\lambda^{-2}$ (see text)  for the  \Y~ and \La~ samples. The full
line is a linear fit to the data. Inset:  pressure dependence of
$T_c$ and linear fit for \Y~ sample}
\label{fig3}
\end{figure}

 Let's now consider the effect of the pressure on
$\lambda^{-2}$ in the low temperature region.  Looking at the
 left panel of Fig. 2, a clear increase of $\lambda^{-2}$ with increasing pressure at low
temperature is visible for \Y. By using the values of
$\lambda^{-2}$ measured at 7 K, we calculated the relative shift
$\Delta \lambda^{-2} /\lambda^{-2}$ = [$\lambda^{-2}$(p) -
$\lambda^{-2}$(p$_0$)]/$\lambda^{-2}$(p$_0$) and plotted it in
Fig. 3 as a function of pressure. The relative shift increases
linearly and monotonously with pressure with a slope 8(1)
$\%$/GPa. The relative shift between the lowest (0.08 GPa) and the
highest pressure (1.05 GPa) is $\Delta \lambda^{-2} /\lambda^{-2}$
= 8.8(8)$\%$. This value, although smaller than that one found in
 \YB~ \cite{KhasanovY124Pres}, is substantially larger
than that expected for an adiabatic electron-lattice interaction
 in conventional superconductors, such as MgB$_2$
\cite{DiCastroPress}, RbOs$_2$O$_6$ \cite{KhasanovRbOs}, and
YB$_6$ \cite{KhasanovYB6}. The presence of a substantial oxygen
isotope effect on the zero temperature magnetic penetration depth
in \Y~ measured by muon spin rotation \cite{Khasanov04} gives a
strong indication that in this system a remarkable
electron-lattice interaction is present, and non-adiabatic effects
are thus expected. The large variation of $\lambda^{-2}$ with
applied pressure, which induces a lattice modification as the
isotope exchange, provides a further confirmation of the relevant
role played by the lattice in hole doped high temperature cuprate
superconductors.

As to the electron-doped compound \La, in the right panel of Fig.
2, $\lambda^{-2}$ in the low temperature region is shown. No clear
trend of $\lambda^{-2}$(7K) with increasing pressure is seen, the
curves coinciding within the error bar. In Fig. 3, the relative
shift $\Delta \lambda^{-2} /\lambda^{-2}$ measured at 7 K is
plotted as a function of pressure. In contrast to  \Y,  there is
no variation of $\Delta \lambda^{-2} /\lambda^{-2}$ with pressure
within the error bar. This is an important result if compared to
the strong variation
 of $\lambda^{-2}$ found in \Y~
(this work) and \YB~ (Ref. ~\cite{KhasanovY124Pres}).  To give a
more quantitative estimation, let's try to estimate the variation
of $\lambda^{-2}$ with pressure, starting from the $zero-{th}$
approximation of a free electron gas. Since
$\lambda^{-2}(0)\propto \omega_p^2$, where $\omega_p$ is the
plasma frequency, then a free electron gas estimate would give
$d\ln\lambda^{-2}(0)/dp=1/B\simeq  0.85 \%/$GPa, where
$B=-dp/d\ln\Omega\simeq 117 $ GPa  is the bulk modulus
\cite{Shaked} and $\Omega$  the volume of the unit cell.
Therefore, for a variation of pressure of about 0.8 GPa, $\Delta
\lambda^{-2} /\lambda^{-2} \simeq$ 0.68 \%, that is of the order
of the error bar in Fig. 3, and  compatible with the experimental
results. The same calculation applied to \Y~ by using $B$ = 156
GPa \cite{Ye} for a variation of pressure of about 1 GPa would
give $\Delta \lambda^{-2} /\lambda^{-2} \simeq$  0.64 \%, a value
one order of magnitude smaller than the experimental result. It is
clear that the effects of the band structure and of the
electron-phonon interaction should in this case be taken into
account. However, this would imply the knowledge of parameters, as
the pressure dependence of the Fermi energy density of state and
of the electron-phonon coupling, which are not unambiguously
determined in the case of HTS's and in particular for \Y. Beside
these considerations, it is worth to recall that most of the
strong effect of pressure on $\lambda^{-2}$ found in \YB~
\cite{KhasanovY124Pres} was ascribed to the change of the
effective mass, or, in other words, to the pressure dependence of
the non-adiabatic electron-phonon coupling, thus supporting our
conclusions about the \Y~ results.

From these considerations, one can argue that the absence of a
pressure effect in the electron-doped ILS \La~ can be ascribed to
the negligible role played by non-adiabatic effects in this
system. To reinforce this guess, we recall that, in a
superconductor close to the clean limit, the zero-temperature
superfluid density is essentially determined by
$\lambda^{-2}(0)\propto$ $n_s/m^{\ast}$
\cite{Uemura,Bernhard,Zhao97}, where $n_s$ is the superconducting
charge carrier density and $m^{\ast}$ is the effective mass of the
superconducting carriers. Therefore, a variation of
$\lambda^{-2}(0)$ can be ascribed either to a change of $n_s$ or
to a change of $m^{\ast}$ or both \cite{Uemura,Bernhard,Zhao97}.
In this respect, the results obtained on the \La~ indicate that
either both $n_s$ and $m^{\ast}$ do not vary with pressure, or
both vary of an identical relative amount. We think that the
second hypothesis is highly unlikely. Indeed, an important hint is
given by the zero pressure effect on $T_c$, which indicates that
the change in $n_s$, due to possible pressure induced charge
transfer, cannot be substantial, as mentioned above. On the other
hand, recent studies of the temperature dependence of the
penetration depth in \La~ indicate the presence of a preponderant
s-wave component in the symmetry of the superconducting order
parameter \cite{KhasanovILS,ChenPRL02,White,Liu}. This suggests
for \La~  a behavior more similar to conventional BCS
superconductors, where pressure has been shown to have no effect
on the penetration depth
\cite{KhasanovRbOs,KhasanovYB6,DiCastroPress}.

This consideration supports other experimental findings which
suggest similarities between \La~ and conventional
superconductors. For example, it was found \cite{Jung} that $T_c$
in \La~ is much more affected by magnetic
 impurities (Ni) than by non-magnetic ones (Zn), as  observed in conventional
superconductors. Moreover, bulk and surface sensitive techniques
show absence of pseudogap \cite{Liu,ChenPRL02} in this $n$-HTS.

In conclusion, we performed a comparative study of the pressure
effects on the magnetic penetration depth in a $p$-HTS (\Y) and a
$n$-HTS (\La), by means of magnetization measurements. The results
for  \Y~  confirm the non-adiabatic character of the
electron-lattice interaction in the $p$-HTS's. On the contrary,
the $n$-HTS \La~ shows absence of non-adiabatic effects, as found
in conventional BCS superconductors like RbOs$_2$O$_6$, YB$_6$,
and MgB$_2$. Our results, together with the previously obtained
indications of the presence of an s-wave symmetry order parameter
in  this system, strongly suggest that there are fundamental
differences in the electronic properties of  $p$-HTS's and the
$n$-HTS ILS superconductor \La. The present work open to future
pressure experiments on other electron doped HTS's, in order to
check if the adiabatic behavior is a universal property of
$n$-HTS's, looking for further evidences for the asymmetric
consequences of hole- and electron-doping in cuprate
superconductors.

The authors gratefully acknowledge C. Grimaldi and P. Postorino
for  helpful discussions on the subject. This work was partly
supported by the EU Project CoMePhS,  the Swiss National Science
Foundation SCOPES grant No. IB7420-110784, and the K. Alex
M\"uller Foundation.

%*****************************************************************************************
%\newpage

%
\end{document}